\documentclass[apj,twocolumn]{openjournal}
\usepackage{amsmath}
\usepackage{booktabs}
\usepackage{multirow}
\usepackage{color}
\usepackage{soul}

\usepackage{siunitx}

\DeclareSIUnit{\Msun}{M_\odot}

\newcommand{\spectral}[3]{#1\,{\sc #2}\,{$\lambda #3$}}

\usepackage[dvipsnames]{xcolor} 

\usepackage[breaklinks,colorlinks,citecolor=blue,urlcolor=blue]{hyperref}

\defcitealias{Shen2018}{S18}

\usepackage{orcidlink}

\renewcommand{\arraystretch}{1.2}  
\usepackage{listings}
\usepackage{color}
\definecolor{dkgreen}{rgb}{0,0.6,0}
\definecolor{gray}{rgb}{0.5,0.5,0.5}
\definecolor{mauve}{rgb}{0.58,0,0.82}
\definecolor{golden}{rgb}{0.86,0.65,0.01}
\begin{document}


\title{Complex spectral variability and hints of a luminous companion in the Be star + black hole binary candidate ALS 8814}

\author{\vspace{-30pt}Kareem El-Badry\,\orcidlink{0000-0002-6871-1752}$^{1,2}$}
\author{Matthias Fabry\,\orcidlink{0000-0003-4200-7852}$^{3}$}
\author{Hugues Sana\,\orcidlink{0000-0001-6656-4130}$^{4,5}$}
\author{Tomer Shenar\,\orcidlink{0000-0003-0642-8107}$^{6}$}
\author{Rhys Seeburger\,\orcidlink{0000-0001-8898-9463}$^{2}$}

\affiliation{$^1$Department of Astronomy, California Institute of Technology, 1200 E. California Blvd., Pasadena, CA 91125, USA}
\affiliation{$^2$Max Planck Institute for Astronomy, Königstuhl 17, 69117 Heidelberg, Germany}
\affiliation{$^3$Department of Astrophysics and Planetary Science, Villanova University, Villanova, PA 19085, USA}
\affiliation{$^4$Institute of Astronomy, KU Leuven, Celestijnenlaan 200D, 3001 Leuven, Belgium}
\affiliation{$^5$Leuven Gravity Institute, KU Leuven, Celestijnenlaan 200D, box 2415, 3001 Leuven, Belgium}
\affiliation{$^6$School of Physics and Astronomy, Tel Aviv University, Tel Aviv 6997801, Israel}
\email{Corresponding author: kelbadry@caltech.edu}

\begin{abstract}
The emission-line binary ALS 8814 was recently proposed as a Be star + black hole (BH) binary based on large-amplitude radial velocity (RV) variations of a Be star and non-detection of spectroscopic features from a luminous companion. We reanalyze low- and medium-resolution LAMOST spectra of ALS 8814 and show that the system's spectroscopic variability is considerably more complex than previously recognized. Inspection of the system's trailed spectra reveals the presence of a second set of absorption lines that move in anti-phase with the Be star. In addition, the emission and absorption lines exhibit inconsistent RV variability, suggesting that they trace different stars. Using spectral disentangling, we recover the spectrum of a rapidly rotating companion whose broad, shallow lines were previously undetected. Time-variability in the emission lines complicates interpretation of the disentangled spectrum, such that the physical parameters of the components are still uncertain, but we find with simulations that emission line variability alone is unlikely to explain all signatures of the companion. The system's high {\it Gaia} \texttt{RUWE} value suggests a third luminous companion, making ALS 8814 a likely hierarchical triple. Although it is unlikely to contain a BH, the system is unusual, with the largest RV semi-amplitude observed in any known classical Be star and a companion that does not appear to be stripped. More extensive spectroscopic monitoring and high-resolution imaging will be necessary to fully characterize the system's orbital architecture, stellar parameters, and evolutionary status.

\end{abstract}

\maketitle

\section{Introduction}
Detached binaries containing a massive OB star and a black hole (OB+BH binaries) are predicted to be relatively common, representing $\sim 3\%$ of all massive stars \citep[e.g.][]{langerPropertiesOBStarblack2020}. Such systems are detectable progenitors of gravitational wave sources and BH high-mass X-ray binaries, and constraints on their population demographics can constrain key uncertainties in binary evolution models \citep[e.g.][]{Janssens_gaia_2022, Tauris2023, Wang2024, Xu2025}. Spectroscopic searches for OB+BH binaries have thus far yielded two solid detections \citep{shenarXrayquietBlackHole2022, mahyIdentifyingQuiescentCompact2022}, and efforts to find additional systems are ongoing.

One predicted class of OB+BH binaries are Be star+BH binaries \citep[e.g.][]{Raguzova1999, Zhang2004, Shao2014,  Grudzinska2015, Rocha2024}. Be stars are rapidly rotating main-sequence stars with decretion disks which give rise to emission-line spectra \citep{Rivinius2013, Rivinius2024b}. The origin of Be stars is an area of active research, but there is strong evidence that a significant fraction of Be stars are spun-up accretors in post-interaction binaries \citep[e.g.][]{Pols1991, Shao2014, shenarHiddenCompanionLB12020, bodensteinerHR6819Triple2020, Wang2021, el-badryStrippedcompanionOriginBe2021, El-Badry2022, Klement2024, Rivinius2025}. Be+BH binaries may thus be a natural outcome of mass transfer in massive binaries, but their predicted occurrence rate is sensitive to several poorly constrained ingredients in binary evolution models \citep[e.g.][]{Belczynski2009}.

About 200 Be+neutron star binaries have been discovered in the Milky Way and Magellanic Clouds, mostly via their X-ray outbursts \citep[e.g.][]{Reig2011, Coe2015, Fortin2023}, but no secure Be+BH binaries are known. 
 One candidate, MWC 656, was reported by \citet{Casares2014}. However, recent work has shown that the companion to the Be star is likely a helium star or white dwarf, not a BH \citep{Rivinius2024, janssensMWC656Unlikely2023}. Additional BH candidates associated with Be stars have been reported over the last several years \citep[e.g.][]{liuWideStarblackholeBinary2019, Rivinius2020, Lennon2022}, but these systems have since been reinterpreted to not contain BHs \citep{abdul-masihSignature70solarmassBlack2020,shenarHiddenCompanionLB12020, bodensteinerHR6819Triple2020, el-badryStrippedcompanionOriginBe2021, frostHR6819Binary2022, El-Badry2022_ngc2004, Muller-Horn2025}. In several cases, the binaries were found to be post-interaction systems containing luminous companions rendered difficult to detect due to rapid rotation. 
 
Recently, \citet{An2025} reported discovery of a new Be+BH binary candidate, ALS 8814, based on radial velocity measurements from the LAMOST time-domain spectroscopic survey \citep{Cui2012,Wang2021_lamost}. ALS 8814 is a bright ($V= 10.2$) source in the Galactic plane at a distance of order 1 kpc. \citet{An2025} report strong emission lines, characteristic of a Be star, which trace an orbit of $P_{\rm orb} = 176.55$ d and a mass function $f(M) = 2.16\pm 0.17\,M_{\odot}$. Unlike in the BH impostors containing Be stars discussed above,  it is the Be star that displays large-amplitude RV variability in ALS 8814, not the companion.

Adopting a Be star mass of $M_{\rm Be} \approx 11\,M_{\odot}$, \citet{An2025} infer a dynamical minimum companion mass of $M_2 \gtrsim  10\,M_{\odot}$. Since they do not detect spectroscopic features from a luminous companion, they surmise that the companion is a BH. They also infer an inclination of $i \approx 36$ deg from modeling of the Be star's emission line profiles, and thus infer a BH mass of $M_2 \approx 23.5\,M_{\odot}$. These conclusions rest critically on the assumption that the system is single-lined.

Here, we reanalyze the archival data of ALS 8814. We present evidence for phase-locked, spectroscopic line profile variability that is consistent with the presence of a second luminous star. We further show that the emission and absorption lines appear not to trace the same object. Using spectral disentangling -- a method to uncover hard-to-detect companions from multi-epoch spectra \citep[e.g.][]{simonDisentanglingCompositeSpectra1994} -- we recover the spectrum of a rapidly rotating, early-type companion whose shallow and broadened lines eluded detection in previous analyses. Our results imply that ALS 8814 is probably not a Be star + BH binary, but rather a double-lined system composed of at least two luminous stars. We also discuss evidence that the system likely contains a third luminous component that has not yet been resolved. 

The remainder of this paper is organized as follows. In Section~\ref{sec:data}, we describe the data to be analyzed, including low- and medium-resolution spectra, light curves from the TESS and K2 missions, and {\it Gaia} astrometry. Section~\ref{sec:analysis} is focused on analysis of the spectra, highlighting evidence for a luminous companion and results of spectral disentangling. Astrometry and possible evidence for a third component in the system are the subjects of Section~\ref{sec:ruwe}. We discuss the nature and possible formation history of the system in Section~\ref{sec:discussion} and conclude in Section~\ref{sec:conclusions}.

\section{Data}
\label{sec:data}

\subsection{Medium-resolution spectra}
We retrieved 26 spectra of ALS 8814 from the LAMOST medium-resolution survey (MRS) published in LAMOST DR10. These are the same spectra analyzed by \citet{An2025}, except that we do not analyze four additional spectra studied in that work which are not available through the LAMOST DR10 web interface.  

The spectra have resolution $R\approx 7,500$ and cover the wavelength ranges of $\approx 4950-5300$\,\AA\, and $\approx 6300-6800$\,\AA. For B stars, the strongest features in this wavelength range are H$\alpha$ and three relatively weak He lines with air wavelengths of 5015.67, 5047.74, and 6678.15\,\AA. A general challenge for the analysis is that the MRS spectra do not contain most of the strong He lines or higher-order Balmer lines (which are less contaminated by emission than H$\alpha$) that are typically used to classify hot stars. Some of the blue arm spectra extend as far blue as 4900\,\AA\,\,and cover \spectral{He}{I}{4921.93}; we also analyze this line when possible.

The typical peak SNR is $\approx 200$ per pixel in both arms. The spectra are barycenter-corrected and reported in vacuum wavelength. Each spectrum represents a coadd of multiple exposures taken in a single night, but we limit our analysis to the coadds. We refer to \citet{Liu2020} for a summary of the LAMOST MRS data and to \citet{Wang2021_lamost} for a description of the LAMOST Time Domain Survey, which obtained many of the spectra of ALS 8814. 

We normalized the spectra from the blue and red arms separately by fitting a 7th-order polynomial to manually-selected wavelength regions free of detectable absorption and emission lines. We chose continuum regions with a typical separation of 20\,\AA\, and inspected all the spectra to verify that the continuum regions are free of strong features or artifacts. 

\subsubsection{Stationary features due to interstellar absorption}
The MRS spectra contain several stationary lines due to interstellar absorption. Beginning with the list of diffuse interstellar bands (DIBs) curated by \citet{Fan2019}, we identified 15 such lines detectable in the spectra of ALS 8814. Since stationary features will bias the results of spectral disentangling and our inference of potential luminous companions, we removed these lines from the data by fitting and subtracting a Gaussian model of each line. We fixed the amplitude, width, and central wavelength of each line across all epochs. For DIBs that are blended with photospheric lines -- the most important of which is at an air wavelength of 6672.23\,\AA, close to \spectral{He}{I}{6678.15} -- we fit DIB parameters in the epoch at which the line is most separated from the photospheric lines. For the relatively weak DIBs in the spectral regions where we perform disentangling, we verified that a Gaussian fit satisfactorily removes the lines and produces residuals free of additional structure. 

\subsubsection{Radial velocities for the primary}
As described by \citet{An2025}, the MRS spectra contain emission lines and relatively narrow absorption lines that shift coherently from epoch to epoch. We attempted to measure radial velocities (RVs) for several of these lines via cross-correlation with a TLUSTY model spectrum \citep{Hubeny1995}  from the BSTAR06 grid \citep{Lanz2007} with $T_{\rm eff} = 26\,{\rm kK}$ and $\log g = 4.0$, the atmospheric parameters inferred by \citet{An2025}.

Our inferred RVs vary somewhat depending on the line being fit but are broadly consistent with the measurements of \citet{An2025}. For consistency with their work, we adopt their orbital solution and ephemeris (their Table 1) in our preliminary analysis here; we refine the solution using  disentangling in Section~\ref{sec:disentangling}. Their solution has  orbital period $P_{\rm orb} = 176.55\pm 0.11$\,d, eccentricity $e = 0.23\pm 0.02$, and RV semi-amplitude $K_1=50.41\pm 1.3\,{\rm km\,s^{-1}}$, corresponding to a companion mass function $f_m=2.16\pm 0.17\,M_{\odot}$.

\subsection{Low-resolution spectra}
We additionally retrieved nine spectra of ALS 8814 from the LAMOST low-resolution survey \citep[LRS;][]{Cui2012, Zhao2012}. These have resolution $R\approx 1800$ and span $(3700-9000)$\,\AA, covering the higher-order Balmer lines and the Paschen series, neither of which are covered by the MRS spectra. These data are also studied by \citet{An2025}, who limit their analysis to the six highest-quality visits. The spectra are not consistently calibrated, with the shape of the continuum varying unphysically from epoch to epoch, so we rescale the spectra in narrow regions centered on individual lines in our analysis. 

\subsection{Parallax, distance, and extinction}
\label{sec:plx}
The {\it Gaia} DR3 parallax of ALS 8814 (DR3 source ID 3425138463244391680) is $\varpi = 0.44\pm 0.29$. However, the source has a large \texttt{RUWE} (Section~\ref{sec:ruwe}), indicating a poor astrometric fit. As demonstrated by \citet{El-Badry2025}, the parallax uncertainty should therefore be inflated, to $\varpi=0.44\pm 0.87$, which is consistent with any distance larger than $\sim 0.5$\,kpc. We discuss the origin of the large \texttt{RUWE} in Section~\ref{sec:ruwe}.

\citet{An2025} infer an absolute magnitude $M_G= -2.49\pm 0.56$ for ALS 8814 based on the strength and morphology of its Balmer decrement, and on this basis infer a distance of $d\approx 1.1\pm 0.2$\,kpc. However, our analysis in this work shows that ALS 8814 likely contains two stars of similar brightness, in addition to a possible tertiary. This casts doubt on the reliability of the $M_G$ inference from the Balmer decrement as well as the resulting constraint on distance. 

We note that \citet{An2025} estimated an extinction $E(B-V) \approx 0.89$ based on the strength of DIBs and the shape of the broadband spectral energy distribution (SED), which is significantly larger than the $E(B-V) \approx 0.31$ predicted at $d=1.1\,{\rm kpc}$ by the \citet{Green2019} 3D dust map or the $E(B-V) \approx 0.34$ predicted by the \citet{Edenhofer2024} 3D dust map. While it is possible that some of the extinction is due to circumstellar material or the Be star's disk, the DIBs are stationary and do not trace the disk's motion. The extinction inferred from DIBs and the SED would be consistent with that predicted by the 3D dust map for a distance of $d\approx 2.5$\,kpc.

We consider the source's distance quite uncertain. Depending on the nature of the component stars and the uncertain flux and extinction contributions from the disk, the source's SED can be reconciled with distances ranging from 1 to 3 kpc. 

\subsection{Light curves}
\label{sec:lightcurve}
ALS 8814 was observed by both {\it K2} \citep{Howell2014} and {\it TESS} \citep{Ricker2015}. It was classified as a slowly pulsating B star (SPB) by \citet{Armstrong2015} on the basis of its {\it K2} light curve. We extracted the {\it TESS} light curves from sectors 43, 44, and 45 using \texttt{eleanor} \citep{Feinstein2019}. The shape of the power spectrum is typical of Be stars \citep[e.g.][]{Labadie-Bartz2022}, with a broad peak resulting from variability on several closely-spaced frequencies. The strongest variability period in all sectors is $\approx 0.3$ d, with a typical amplitude of 0.5\%. This variability can plausibly be attributed to either $\beta$ Cephei or SPB-type pulsations \citep[e.g.][]{Shi2024} and is unlikely to reflect a rotational or orbital period in the system, being comparable to the dynamical time for an early B star. Besides binary motion and variability in the disk, pulsations are another possible source of line profile variations from epoch to epoch.

\section{Analysis}
\label{sec:analysis}

\begin{figure*}
    \centering
    \includegraphics[width=\textwidth]{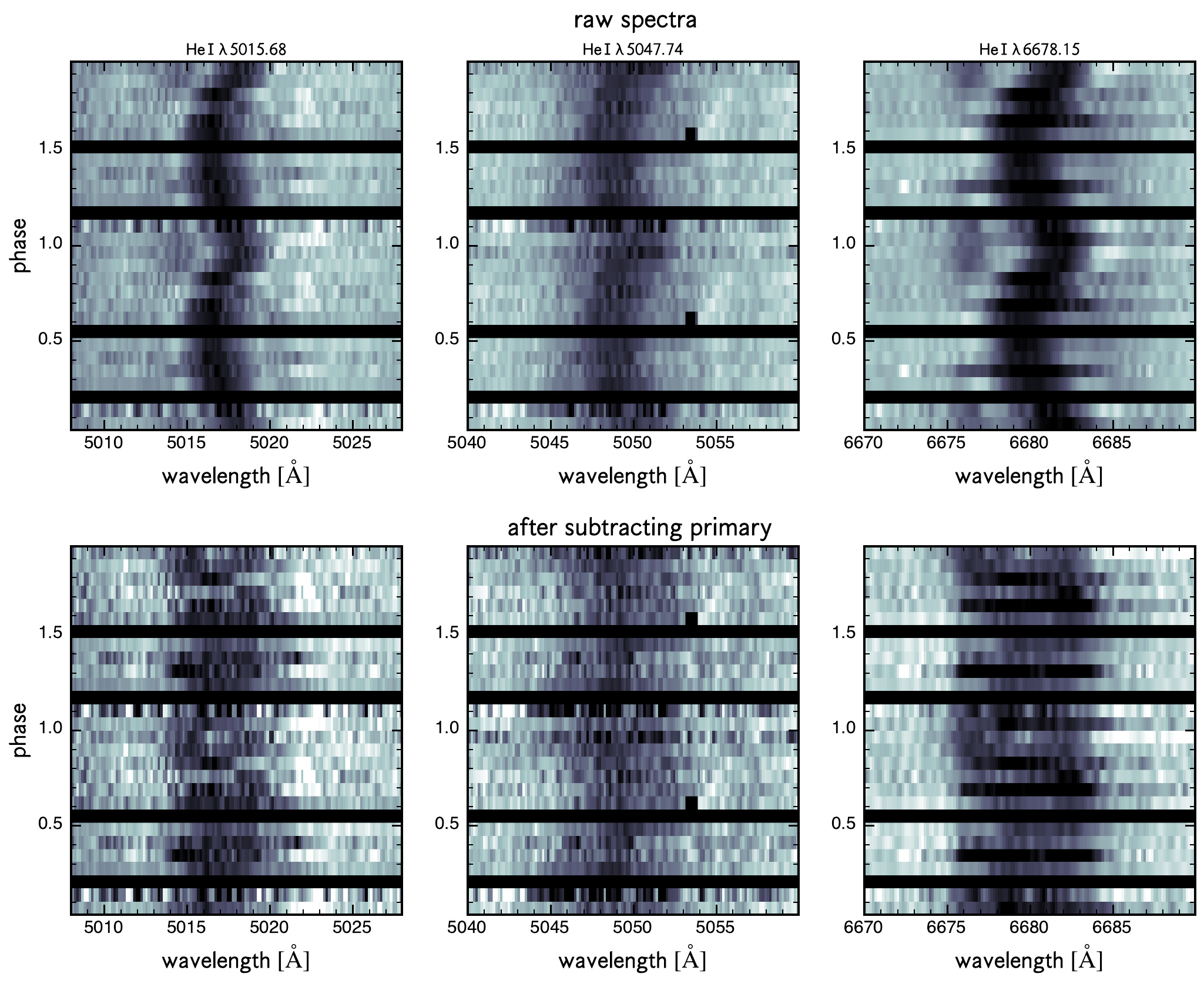}
    \caption{Phase-folded trailed LAMOST MRS spectra of ALS 8814. Top panels show the raw data; bottom panels show data after subtraction of the disentangled mean spectrum of the primary. In the top panels, narrow absorption lines flanked by double-peaked emission lines shift coherently. Traces of broader absorption lines moving in anti-phase can also be discerned; these become more obvious in the bottom panels.  }
    \label{fig:trailed}
\end{figure*}

\subsection{Trailed spectra}

Figure~\ref{fig:trailed} shows trailed normalized spectra of ALS 8814, ordered by orbital phase. The three vertical columns highlight the three strongest He lines that are covered by all the MRS spectra. We average spectra in 15 phase bins and duplicate them once. Here the phase is defined as $\phi = \left[( \mathrm{JD} - T_\mathrm{p} ) \bmod P_{\mathrm{orb}}\right ]/ {P_{\mathrm{orb}}}
$, where $P_{\rm orb} = 176.55$ d is the orbital period and $T_\mathrm{p}=2458042.29$ is the periastron time. 

The top row shows the raw normalized spectra, while the bottom row shows the results of subtracting the disentangled spectrum of the primary at the appropriate RV (section~\ref{sec:disentangling}). In the top row, all three panels show an absorption line moving coherently. However, the \spectral{He}{I}{5015.68} and \spectral{He}{I}{6678.15} lines also show double-peaked emission on both sides of the absorption lines, moving in phase with the absorption. The emission is relatively stable over the course of the orbit, though hints of V/R variations (i.e., changes in the relative amplitude of the red and blue emission peaks) are visible for the \spectral{He}{I}{6678.15} line. Such emission features are common in Be stars and, in ALS 8814, are likely to trace a disk surrounding the primary. 

From the top panel of Figure~\ref{fig:trailed}, we can also discern hints of another set of broader and shallower absorption lines underneath the features tracing the primary. For \spectral{He}{I}{5047.74} and \spectral{He}{I}{6678.15}, a shift of these lines in the opposite direction of the primary is visible even in the raw data: they are maximally blueshifted at phase 1 and maximally redshifted at phase 0.5, suggesting that another luminous object is moving in the opposite direction of the primary. 

To separate the spectra of the two  components, we tested three different spectral disentangling approaches (Section~\ref{sec:disentangling}), which yielded broadly consistent results. In the bottom panels of Figure~\ref{fig:trailed}, we subtract the recovered spectrum of the primary -- shifted to the appropriate RV -- from each observed spectrum. The anti-phase motion of the secondary is now obvious. We have verified that this conclusion is not sensitive to the velocity amplitude of the secondary assumed in disentangling: anti-phase motion is detectable in the residuals even when the primary spectrum is disentangled while assuming a static companion. The lines in the bottom panel of Figure~\ref{fig:trailed} are broader than those of the primary; as we show in Section~\ref{sec:params}, this likely reflects its rapid rotation and the fact that the primary's absorption lines are filled in by disk emission. The fact that the inferred line profiles of the secondary as manifest in the residuals are relatively static disfavors scenarios in which the companion is itself a tight binary. 

\subsection{Different behavior of emission and absorption lines}

\begin{figure*}
    \centering    \includegraphics[width=\textwidth]{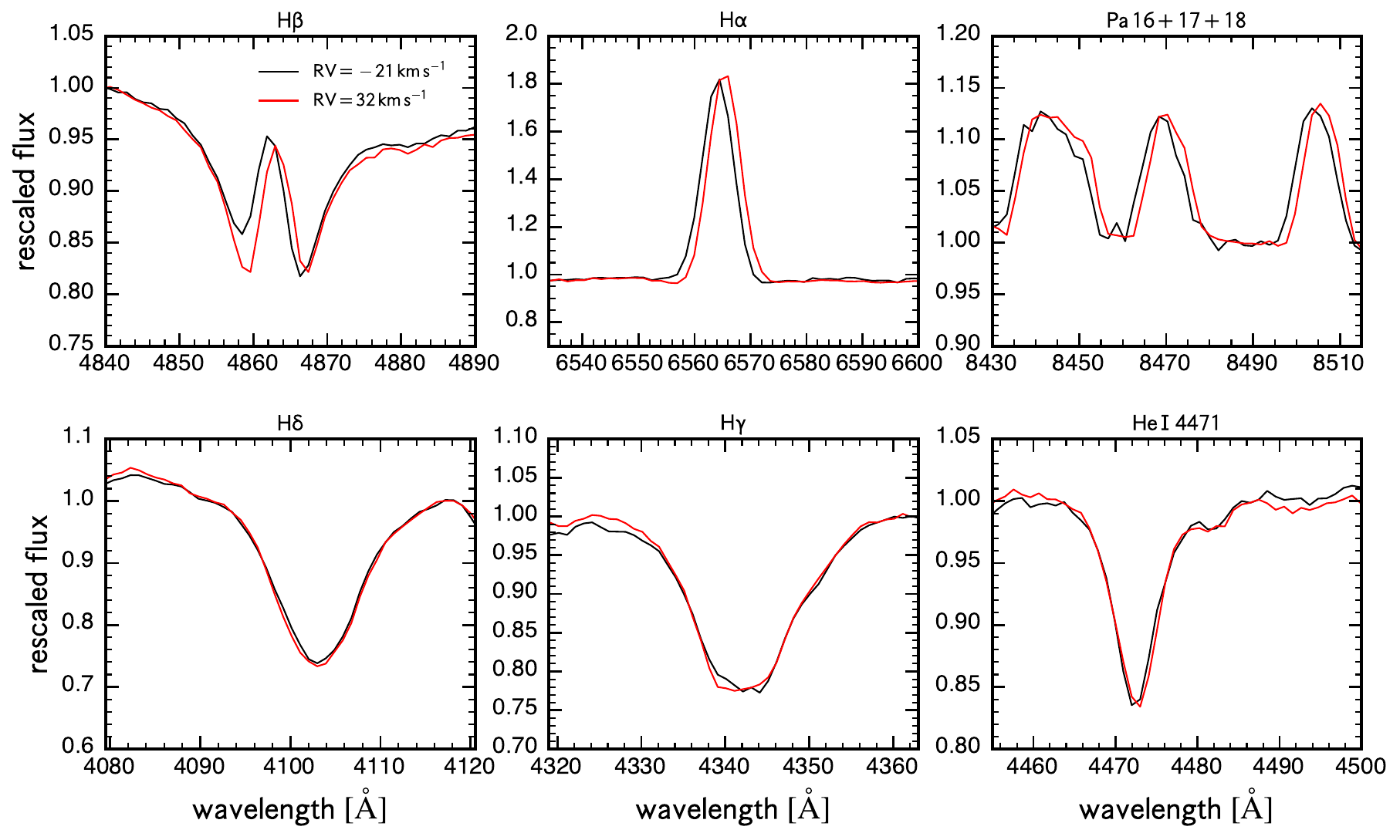}
    \caption{Cutouts of LAMOST low-resolution spectra of ALS 8814 near opposite RV extrema. All panels show a wavelength range corresponding to a fixed velocity range of $\pm 1500\,{\rm km\,s^{-1}}$. Top panels show regions containing emission lines, while bottom panels show strong absorption lines. The emission lines are clearly RV variable, shifting by more than $50\,{\rm km\,s^{-1}}$ between the two epochs. However, no coherent RV variability is visible in the absorption lines. This implies that emission and absorption lines are not tracing the same object. }
    \label{fig:lowres}
\end{figure*}

To probe spectral variability over a broader wavelength range, we investigated the LAMOST LRS spectra taken at a range of phases. Figure~\ref{fig:lowres} compares two spectra taken near the RV extrema of the dataset. Because the calibration of the spectra is inconsistent across epochs, we rescale the spectra in each panel to a consistent flux value. In the low-resolution spectra, the weak He I lines shown in Figure~\ref{fig:trailed} are blended and barely detectable, but the Balmer lines and several stronger He I lines are visible. 

The emission lines in the top row of Figure~\ref{fig:lowres} shift significantly across epochs, which is not surprising given that the RV ephemeris for the primary predicts an RV difference of $53\,\,{\rm km\,s^{-1}}$ between them. The strong absorption lines in the bottom row, however, do not shift coherently with the emission lines: the Balmer absorption lines appear basically stationary across epochs, while the He\,I line shifts only in its core, and still less than the emission lines. This implies that the emission and absorption lines do not trace the same physical object -- at least not completely. We can rule out a scenario where the apparent lack of RV shifts in the absorption lines is due to problems with the wavelength solution that by coincidence counteract true RV variability, since the top left panel of Figure~\ref{fig:lowres} shows that within the H$\beta$ line, the emission core clearly shifts while the absorption wings are approximately static.

The most likely explanation is that the strong absorption lines contain roughly equal flux contributions from two luminous sources moving in opposite directions, such that their contributions to the mean RV of the line approximately cancel \citep[e.g.][]{sanaNonthermalRadioEmitter2011}. This occurs to a lesser extent for the weak lines highlighted Figure~\ref{fig:trailed}, because their cores are more strongly affected by emission lines. As a result, the RV shifts of the disky component are much more obvious than those of the companion. 

\begin{figure*}
    \centering    \includegraphics[width=\textwidth]{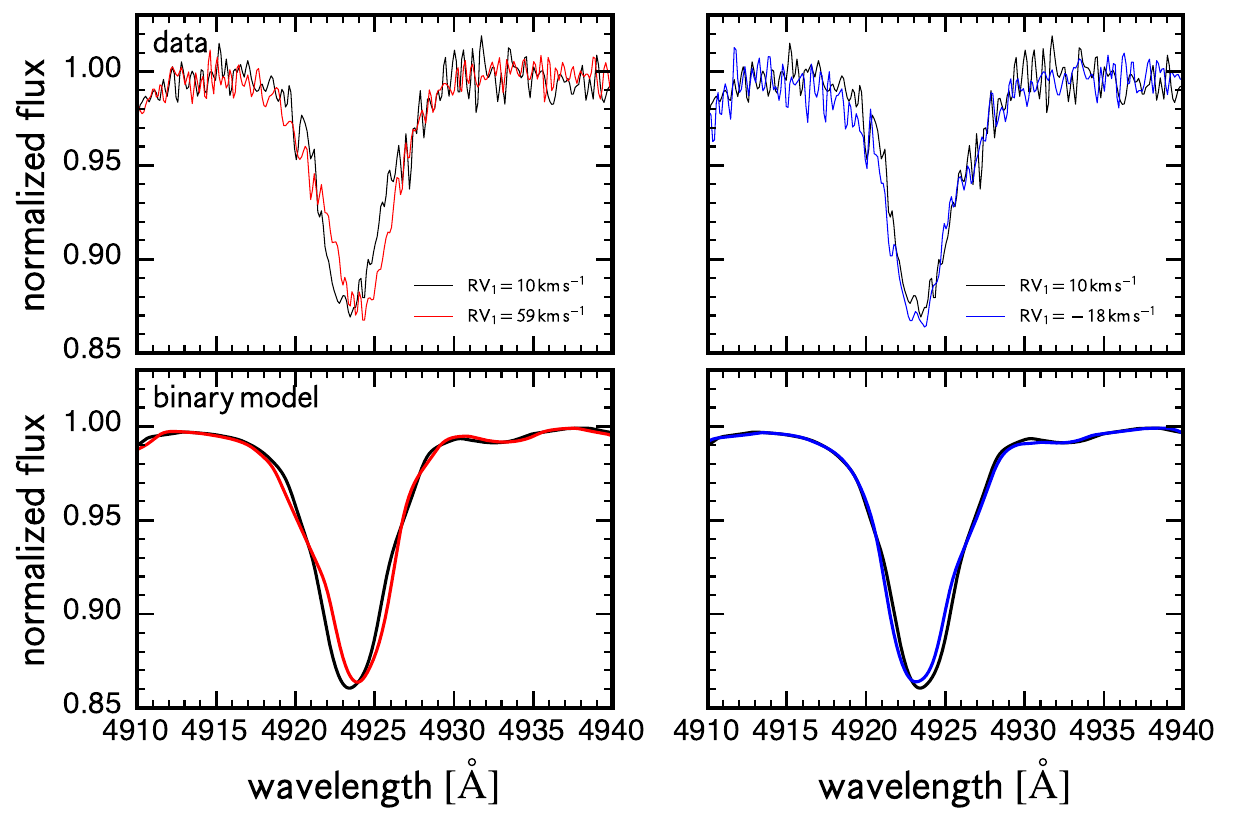}
    \caption{Variability of \spectral{He}{I}{4921.93}, the strongest He line covered by the MRS data. Top panel compares the observed MRS data between a visit near opposition (black) and visits near opposite conjunctions (red and blue). The RVs listed in each panel are those predicted by the ephemeris of the Be star. Bottom panel shows a simulation with two luminous stars (see text for details). The observed line cores shift significantly between epochs, while the wings are nearly stationary. This behavior is reproduced by the simulated spectra. In the simulation, it occurs because the line cores of the Be star -- which are narrow, perhaps due to infilling by disk emission -- dominate near line center, while both components contribute in the wings. }
    \label{fig:medres_HI}
\end{figure*}

To better understand the variability of the He\,I lines, we inspected the MRS spectra in the vicinity of the \spectral{He}{I}{4921.93} line, which is the strongest He line covered by the MRS data. Because the line is near the blue edge of the MRS wavelength coverage, it is included in the published spectra only in about half of the MRS visits. In Figure~\ref{fig:medres_HI}, we compare a spectrum taken near opposition to one observed near the Be star's RV maximum (red) and another observed near its RV minimum (blue). The core of the line shifts significantly between epochs, with a shift comparable to that predicted by the RV ephemeris of the Be star. However, the wings of the line shift less, and in the case of the most redshifted visit, actually shift in the opposite direction of the wings.

The bottom panels of Figure~\ref{fig:medres_HI} show a simulation in which we model two equally-bright stars moving in anti-phase with equal RV amplitudes. We assume $T_{\rm eff} = 26\,{\rm kK}$ and $\log g = 4$ for both components, $v\sin i =130\,{\rm km\,s^{-1}}$ for the primary, and $v\sin i=250\,{\rm km\,s^{-1}}$ for the secondary. These values are motivated by the results of our spectral disentangling (Section~\ref{sec:disentangling}). As we discuss in Section~\ref{sec:params}, the true rotation velocity of the Be star primary is likely higher than $130\,{\rm km\,s^{-1}}$, but this value produces a spectrum that approximates the shape of the combined stellar and disk spectrum. 

The simulated spectra reproduce the behavior of the observed data, including both the shift of the line cores and the near-stationarity of the wings. We consider other possible models that could explain the variability of the absorption lines in Section~\ref{sec:lum_tertiary}.

\subsection{Spectral disentangling}
\label{sec:disentangling}

To separate the spectra of the two luminous components suggested by Figures~\ref{fig:trailed} and~\ref{fig:lowres}, we performed spectral disentangling on the MRS spectra. Spectral disentangling operates under the ansatz that all the observed spectra are produced by summing together the time-invariant spectra of two components, with known Doppler shifts between epochs. Disentangling does not uniquely determine the flux ratio between the two components, which must be constrained by comparison of the disentangled spectra to models.

Our fiducial disentangling is based on the  wavelength-space disentangling method of \citet{simonDisentanglingCompositeSpectra1994}. We used the implementation of the algorithm described by \citet{Seeburger2024} without regularization using template spectra and assume a flux ratio of 0.5, meaning that both components contribute equal light in the mid optical. 

Because the ephemeris from \citet{An2025} was derived while neglecting the presence of the second luminous component, we refit the orbit as part of disentangling. We fix the orbital period and center-of-mass velocity to the best-fit values from their solution, but allow the RV semi-amplitudes of the primary and secondary, $K_1$ and $K_2$, as well as the eccentricity, $e$, periastron time, $T_\mathrm{p}$, and argument of periastron, $\omega$, to vary.

We fit all the MRS spectra in a $20$\,\AA-wide region centered on the \spectral{He}{I}{6678.15} line, which is the strongest He line covered by all the MRS data. For each set of ($K_1,K_2,e,T_\mathrm{p}, \omega$), we predict RVs for both components in all epochs, solve the disentangling problem, and calculate the total $\chi^2$, which quantifies how well two disentangled spectra can reproduce all the data. Finally, we use \texttt{emcee} \citep{Foreman-Mackey2013} to sample from the posterior, using a likelihood function $\log \mathcal{L} = -0.5\chi^2$ and broad, flat priors. The results are reported in Table~\ref{tab:ephem}. The solution for the primary is qualitatively similar to that inferred by \citet{An2025}, but the eccentricity, periastron time, and argument of periastron are all moderately discrepant. The best-fit secondary semi-amplitude, $K_2=24.1\pm 1.6\,{\rm km\,s^{-1}}$, implies a mass ratio $M_2/M_1 \approx 2$, but as we discuss below, the best-fit $K_2$ varies with the spectral region being fit, implying that the formal uncertainties on $K_2$ are significantly underestimated.

\begin{table}
\centering
\caption{Orbital ephemeris from disentangling the \spectral{He}{I}{6678.15} line.}
\label{tab:ephem}
\begin{tabular}{@{}p{0.5\columnwidth}r@{}}
\toprule
Orbital period & $P_{\rm orb} = 176.55$\,d (fixed) \\
Center-of-mass RV & $\gamma = 16.99\,{\rm km\,s^{-1}}$ (fixed) \\
Primary semi-amplitude & $K_1 = 51.6 \pm 1.5$\,km\,s$^{-1}$ \\
Secondary semi-amplitude & $K_2 = 24.1 \pm 1.6$\,km\,s$^{-1}$ \\
Eccentricity & $e = 0.31\pm 0.02$ \\
Periastron time & $T_\mathrm{p} = 2458038.5 \pm 1.4$ JD UTC \\
Argument of periastron & $-0.23\pm 0.05\,{\rm rad}$ \\
\bottomrule
\end{tabular}
\end{table}

\begin{figure*}
    \centering
    \includegraphics[width=\textwidth]{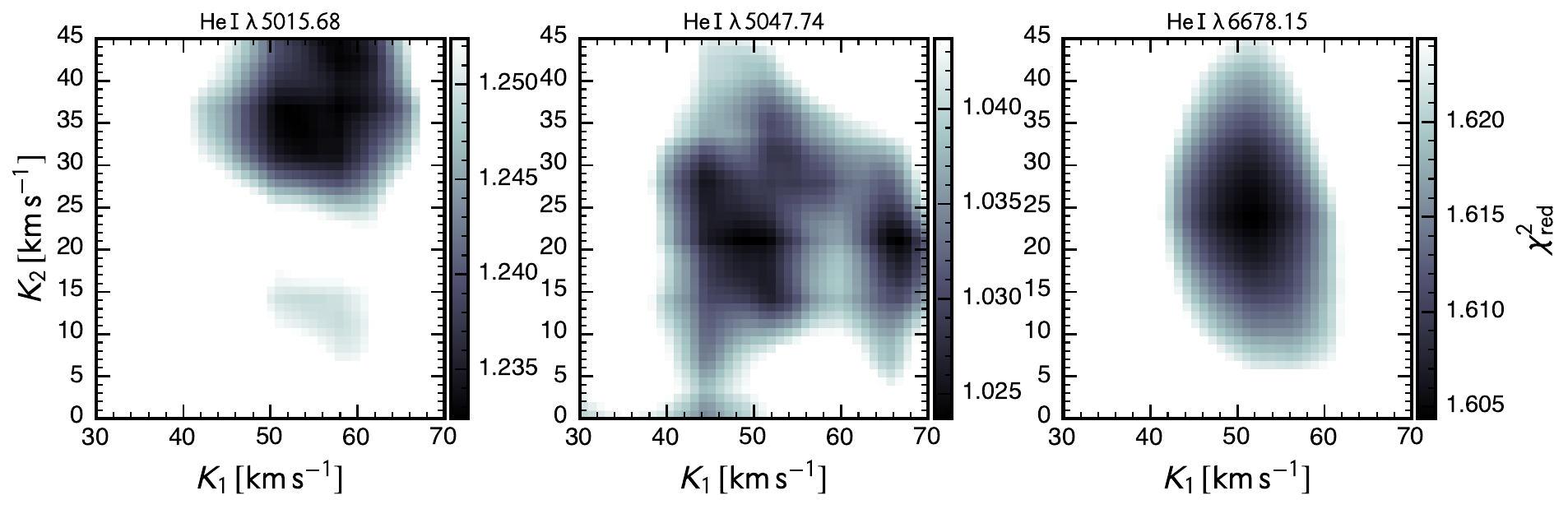}
    \caption{Constraints on the RV semi-amplitude of the primary ($K_1$) and secondary ($K_2$) from wavelength-space disentangling. The three panels show results from the three He I lines highlighted in Figure~\ref{fig:trailed}. Dark shading shows the regions of parameter space for which two components moving in anti-phase can best reproduce all of the observed spectra. $K_1$ is constrained to $51\pm 2\,\rm km\,s^{-1}$ by all three lines, but $K_2$ is more uncertain, with the best-fit value ranging from 22 to 38\,$\rm km\,s^{-1}$.  }
    \label{fig:K1K2}
\end{figure*}

Figure~\ref{fig:K1K2} shows constraints on $K_1$ and $K_2$ for all three He\,I lines covered by the MRS spectra. We fix the orbital parameters other than $K_1$ and $K_2$ to the values reported in Table~\ref{tab:ephem} and plot the total reduced $\chi^2$ across all epochs for each choice of $K_1$ and $K_2$. While the MRS spectra also contain H$\alpha$, obvious changes in the strong emission line profile across epochs make it unsuitable for disentangling.

It is clear from Figure~\ref{fig:K1K2} that $K_1$ is reasonably well-constrained, but $K_2$ is more uncertain: the three different lines are best-fit with $K_2 = 37$, 22, and 24\,$\rm km\,s^{-1}$. All three lines favor $K_2 > 0$, implying antiphase motion of the companion. For each line, the best-fit $K_2$ is also sensitive to the choice of spectral window and how much continuum is included at the edges of the line. All three lines have a best-fit $\chi^2_{\rm red}$ that is larger than 1, meaning that the two disentangled components cannot fully describe all the variance in the observed spectra. This is probably a result of time-variability in the emission lines, but could also be attributed to the presence of a likely third component in the system (Section~\ref{sec:ruwe}).

\begin{figure*}
    \centering
    \includegraphics[width=\textwidth]{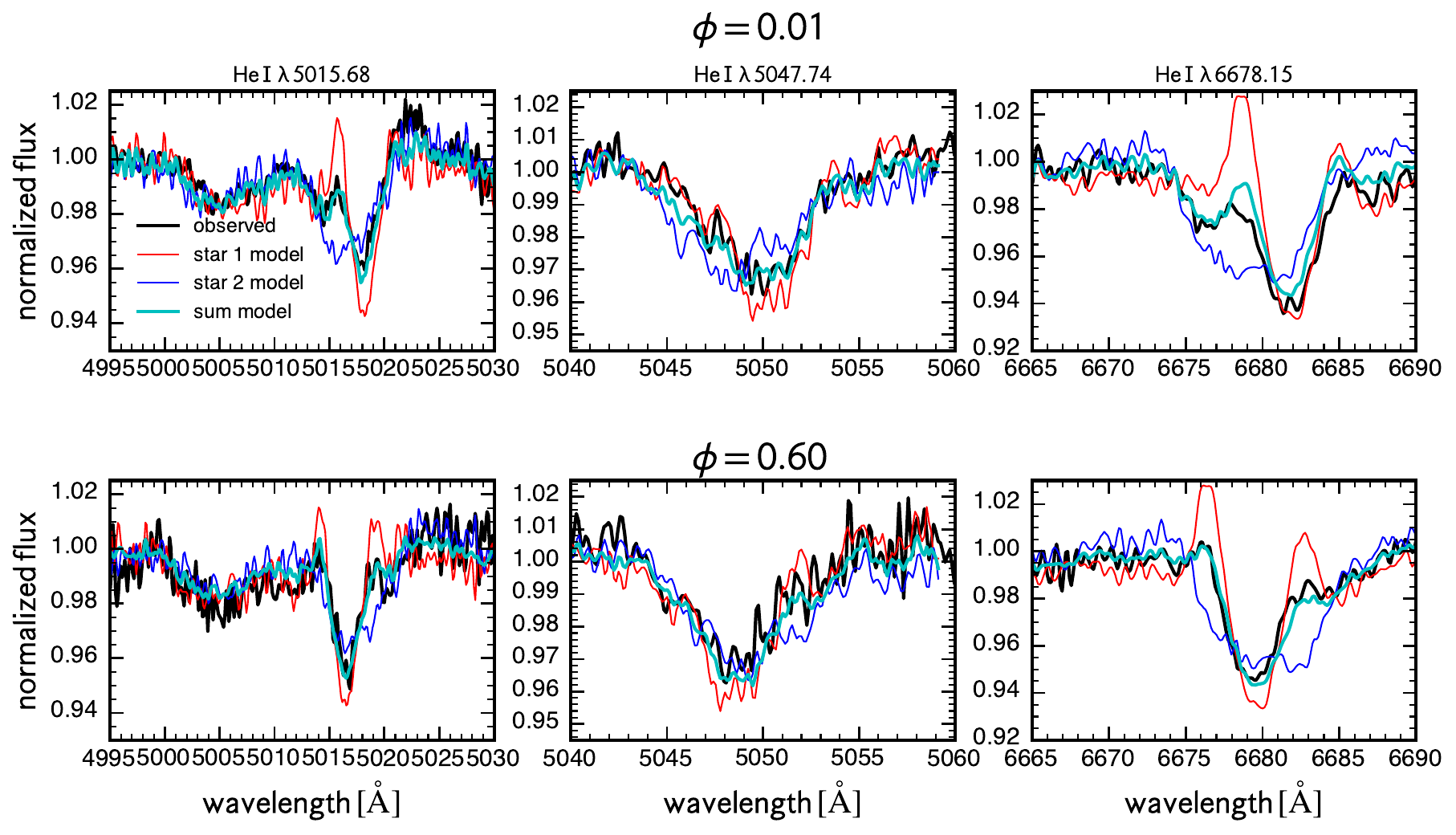}
    \caption{Reconstruction of observed spectra with the disentangled components. Top and bottom rows show epochs near opposite quadratures, while the three columns show three different He I lines. Black lines show observed spectra, while red and blue lines show the disentangled spectra of the primary and secondary. Their sum (cyan) matches the observed spectra well.  }
    \label{fig:disentangled}
\end{figure*}

Figure~\ref{fig:disentangled} shows the reconstruction of the observed spectra with the disentangled models for two spectral epochs chosen to be near opposite quadratures. For all three He\,I lines, there are clear changes in the shape of the observed spectra between epochs. Such changes are not expected for a single luminous component -- at least not without changes in emission lines due to the disk; see Section~\ref{sec:VR_variations} --  but are naturally explained as the sum of two models shifting back and forth in anti-phase. Consistent with the qualitative impression from Figure~\ref{fig:trailed}, the disentangled spectrum of the primary contains double-peaked emission lines on top of absorption lines. 

To validate the disentangling results and assess their sensitivity to the disentangling algorithm, we disentangled the spectra using two additional methods, as described below.

\subsubsection{Fourier spectral disentangling}
Spectral disentangling can also be performed in Fourier space rather than in wavelength space.
\texttt{fd3} \citep{ilijicObtainingNormalisedComponent2004, ilijicFd3SpectralDisentangling2017} determines the disentangled spectra by solving the linear least-squares problem associated to the given orbital parameters and input spectra \citep[][]{simonDisentanglingCompositeSpectra1994, hadravaOrbitalElementsMultiple1995, fabryResolvingDynamicalMass2021}. We disentangled the observed spectra using \texttt{fd3}, assuming the same orbital ephemeris listed in Table~\ref{tab:ephem}. To correct spurious undulations introduced by numerical instability in the disentangling procedure \citep[e.g.][]{Hensberge2008}, we re-normalized the disentangled spectra.

\subsubsection{Shift-and-add}
We also disentangled the same spectra using the shift-and-add algorithm \citep[e.g.,][]{marchenkoWolfRayetBinaryWR1998, Gonzalez2006, mahy2012, shenarShortestperiodWolfRayetBinary2018}.
Shift-and-add is an iterative procedure that uses the disentangled spectra obtained in the $j^\mathrm{th}$ iteration, $A_j$ and $B_j$, to calculate the disentangled spectra for the $j+1^\mathrm{th}$ iteration. The disentangled spectrum of iteration $j$ for the primary (secondary) is produced by subtracting the disentangled spectrum of iteration $j-1$ of the secondary (primary), and co-adding the residuals in the frame of the primary (secondary). The first iteration assumes zero flux for the secondary. For more details, see \citet{shenarHiddenCompanionLB12020}. 


\subsection{Summary of the disentangling results}

\begin{figure*}
    \centering
    \includegraphics[width=\textwidth]{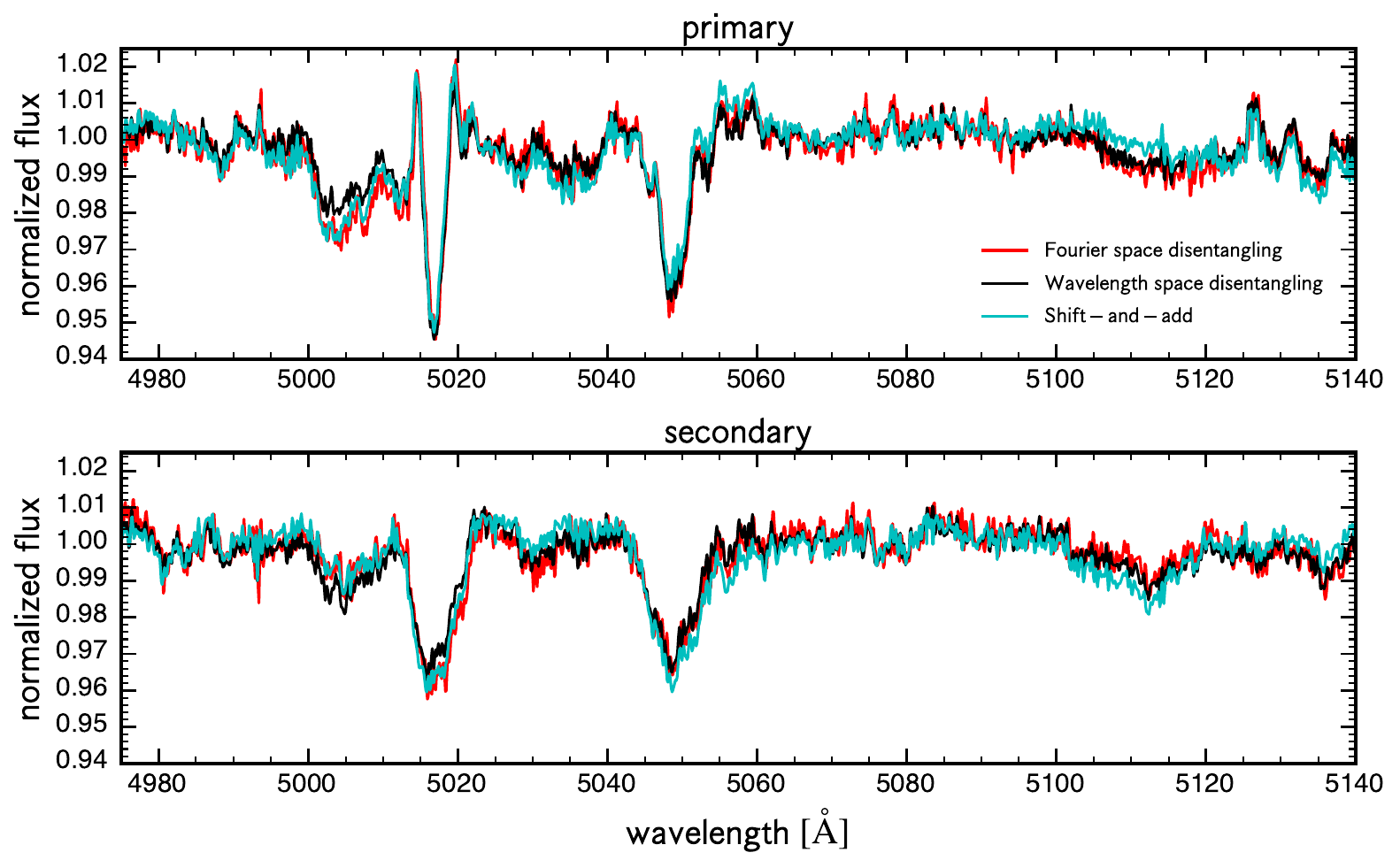}
    \caption{Comparison of the disentangled spectra of the primary (top) and secondary (bottom) produced by three different disentangling methods. Despite minor differences, the three algorithms produce broadly consistent results. The recovered secondary has broad He\,I lines. }
    \label{fig:comparison}
\end{figure*}

Figure~\ref{fig:comparison} compares the disentangled spectra for both components obtained by the three methods. Although the three sets of disentangled spectra are not identical, they all agree that the secondary has clear spectral features in the He I lines. In addition to He\,I lines, the disentangled spectrum of the primary contains several double-peaked Fe II lines in emission. Lines due to C\,II, Si\,II, and N\,II (not all visible in Figure~\ref{fig:comparison}) are also detectable in both emission and absorption.

Although disentangling shows rather unambiguously that another luminous source contributes to the observed spectra, it does not produce a robust measurement of $K_2$ (Figure~\ref{fig:K1K2}).  This is likely simply a consequence of the limited number of spectra and the fact that all the lines covered are rather weak. As we show in Section~\ref{sec:lum_tertiary}, the LRS spectra can also constrain $K_2$. Loose constraints on $K_2$ from disentangling are common in the case of a rapidly rotating companion \citep[e.g.][]{Frost2024}.

\citet{An2025} report that they also performed spectral disentangling on the MRS data and did not find any indications of a luminous companion, which seems inconsistent with our results. The procedure they describe is different from ours in that \citet{An2025} do not actually examine the disentangled spectrum of the secondary. Instead, they recover the disentangled spectrum of the primary (it is unclear what they assumed for $K_2$ in the process), subtract this from each of the individual epoch spectra, and then report that the residuals do not contain clear evidence for a luminous companion. This approach is much less sensitive than inspection of the disentangled spectrum of the secondary, because the latter contains information from all the composite spectra and thus has much higher SNR than any individual visit spectrum.

\citet{An2025} also make arguments against a luminous companion based on the fact that they do not find asymmetries in the H$\alpha$ line that might be expected due to binary motion. However, we find that (a) some epochs do show exactly the type of asymmetries they discuss, and (b) the presence of the Be star's own photospheric absorption line is likely to complicate the total line profile \citep[e.g.][]{El-Badry2020_lb1, Abdul-Masih2020}. The significant variation in  amplitude and width of the H$\alpha$ emission line across epochs make any attempt to infer the companion's properties from its variability quite uncertain.

\label{sec:dis_summary}

\subsection{Parameters of the two stars}
\label{sec:params}

\label{sec:params}
\begin{figure*}
    \centering
    \includegraphics[width=\textwidth]{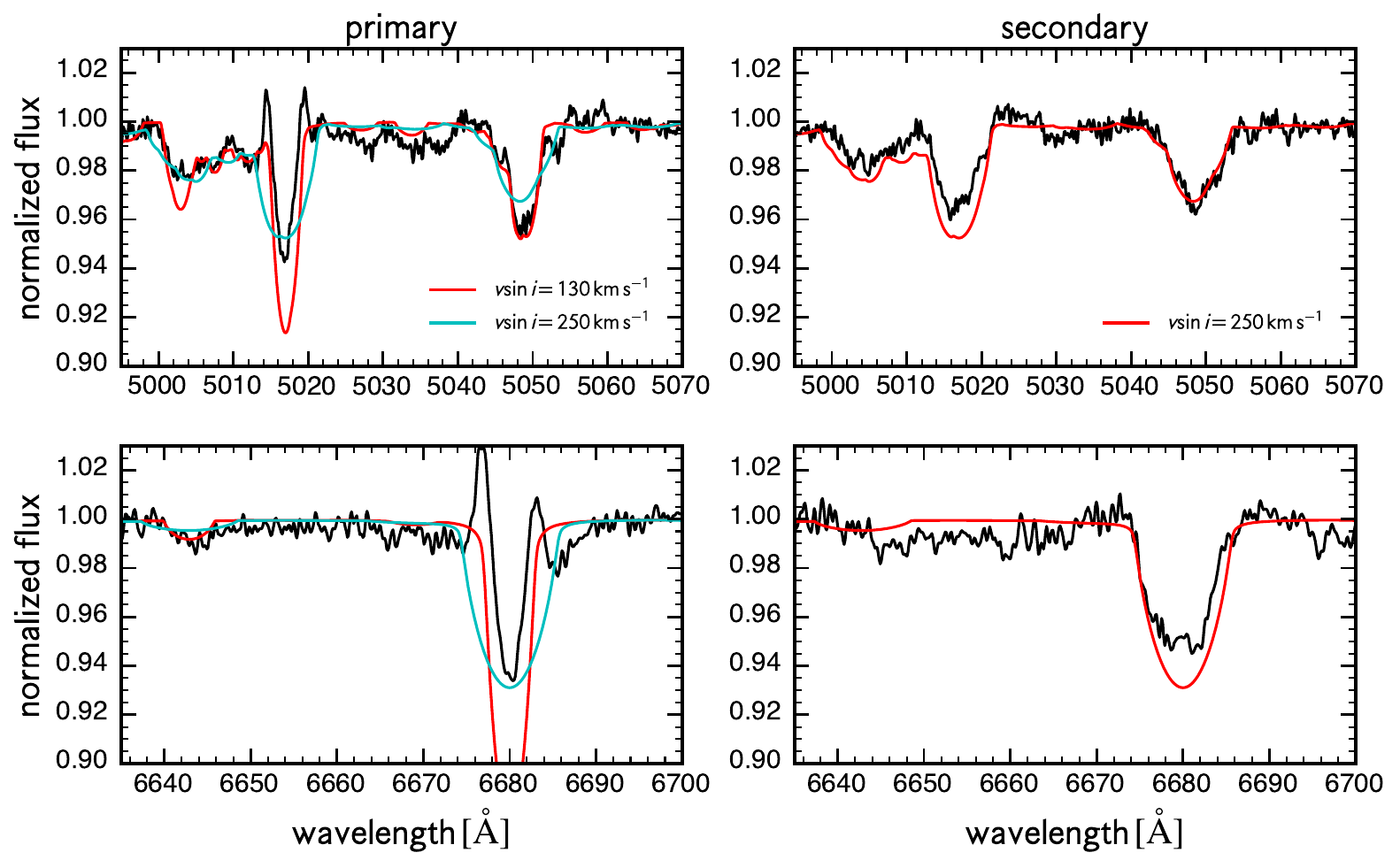}
    \caption{Comparison of the disentangled spectra of the primary and secondary (black lines) to synthetic spectra.  Cyan line in the left panels shows a model with $v\sin i =250\,{\rm km\,s^{-1}}$, as assumed by \citet{An2025}. The data visually appear better fit by a model with $v\sin i =130\,{\rm km\,s^{-1}}$, but this is likely a consequence of infilling from emission lines, and the broad absorption wings of \spectral{He}{I}{6678.15} suggest a value of $v\sin i \approx 250-300\,{\rm km\,s^{-1}}$, which is more reasonable for a Be star. $v\sin i = 250\,{\rm km\,s^{-1}}$ provides a good match for the secondary. Both models have $T_{\rm ef} = 26\,{\rm kK}$ and $\log g =4$. These values represent a reasonable guess for the parameters of the two stars, but given the significant contamination by disk emission, data covering a wider wavelength range are required to robustly measure the atmospheric parameters of both stars. } 
    \label{fig:vsini}
\end{figure*}

The MRS data only allow for disentangling in narrow wavelength ranges containing relatively weak He\,I lines. Together with the fact that the binary's distance is poorly constrained and most lines are contaminated by emission, this makes it infeasible to fully solve for the parameters of both components with the currently available data. Some useful constraints can nevertheless be obtained by considering both the MRS and LRS data:

\begin{itemize}
    \item The LRS data show no He II lines, setting an upper limit on the effective temperatures of both components. In particular, the lack of detected \spectral{He}{II}{4686} features implies $T_{\rm eff} < 30\,{\rm kK}$ for both components. 
    \item The weak RV shifts visible in all the strong absorption lines (Figure~\ref{fig:lowres}) requires a flux ratio near 0.5, such that neither component dominates the flux in any line. Any line dominated by the primary would shift coherently with the emission lines, while any line dominated by the companion would shift in the opposite direction.
    \item The higher-order Balmer lines in the LRS data appear to be minimally contaminated with emission. For B stars, the depth of these lines is sensitive to $T_{\rm eff}$, and the data are best-fit by models with $T_{\rm eff} = 25-30\,{\rm kK}$. Models with lower $T_{\rm eff}$ have lines that are deeper than observed, though it is possible that the observed lines are diluted somewhat by the disk. 
    \item The disentangled spectra of the primary are reasonably well-matched by models with $T_{\rm eff}\approx 26\,{\rm kK}$ and $ \log\left[g/\left({\rm cm\,s^{-2}}\right)\right]\approx4$. Lowering $T_{\rm eff}$ or $\log g$ increases the depth of He\,I and metal lines. However, contamination from the disk significantly complicates interpretation of these spectra. 
\end{itemize}

In Figure~\ref{fig:vsini}, we compare the disentangled spectra to TLUSTY/SYNSPEC models from the BSTAR06 grid \citep{Lanz2007} with $T_{\rm eff}= 26\,{\rm kK}$, $\log \left[ g/ \left({\rm cm\,s^{-2}} \right) \right]=4$, and $[\rm Fe/H]=0$, similar to the values inferred by \citet{An2025} from the shape of the Balmer decrement. We visually adjusted the projected rotation velocity $v\sin i$ to match the width of the He I lines. While it would be preferable to constrain $v\sin i$ using metal lines, these are too weak and contaminated with emission lines for the MRS data to yield a useful constraint.

The disentangled spectrum of the secondary is best-fit by a model with $v\sin i \approx 250\,{\rm km\,s^{-1}}$. The $v \sin i$ of the primary is uncertain, because essentially all of its spectral features are likely contaminated by emission lines from the disk. A low value of $v \sin i = 130\,{\rm km\,s^{-1}}$ best reproduces the cores of the He I lines. However, these are very likely contaminated by double-peaked emission lines, and the broad wings of \spectral{He}{I}{6678.15} suggest a true $v \sin i $ of order $300\,{\rm km\,s^{-1}}$, which is typical for a Be star. However, this measurement is quite uncertain and must be confirmed with higher-quality data. A projected rotation velocity as low as $v \sin i = 130\,{\rm km\,s^{-1}}$ would be difficult to explain astrophysically, since Be stars are expected to rotate at near-critical velocities. We do not attempt to quantitatively fit for the flux ratio and atmospheric parameters of both components here, as a robust constraint will require higher-quality data covering bluer wavelengths.

\section{Evidence for a third component}
\label{sec:ruwe}
ALS 8814 has a high \texttt{RUWE} value in {\it Gaia} DR3: it has \texttt{RUWE} = 18.6, where a good astrometric solution would have \texttt{RUWE} = 1 \citep[e.g.][]{Lindegren_2018_RUWE}. \texttt{RUWE} scales with the residuals of the single-star astrometric solution divided by the epoch level astrometric uncertainties. As a result, it is possible to accurately predict the \texttt{RUWE} value a binary will receive from its orbit, sky position, and apparent magnitude \citep[e.g.][]{El-Badry2025}.

To assess whether the large \texttt{RUWE} of ALS 8814 is plausibly a consequence of the binary's orbital motion, we use \texttt{gaiamock} \citep{El-Badry2024} to predict the system's \texttt{RUWE}. We first assume the companion is dark and adopt the orbital parameters from \citet{An2025}. For an inclination of 36 degrees and a companion mass of $23\,M_{\odot}$, \texttt{gaiamock} predicts \texttt{RUWE}=3-4, depending on the longitude of the ascending node, $\Omega$. Producing a \texttt{RUWE} as large as observed through orbital motion alone would require the companion to be a $\approx 1000\,M_{\odot}$ BH with an inclination of $\approx 7^\circ$.

\begin{figure}
    \centering
    \includegraphics[width=\columnwidth]{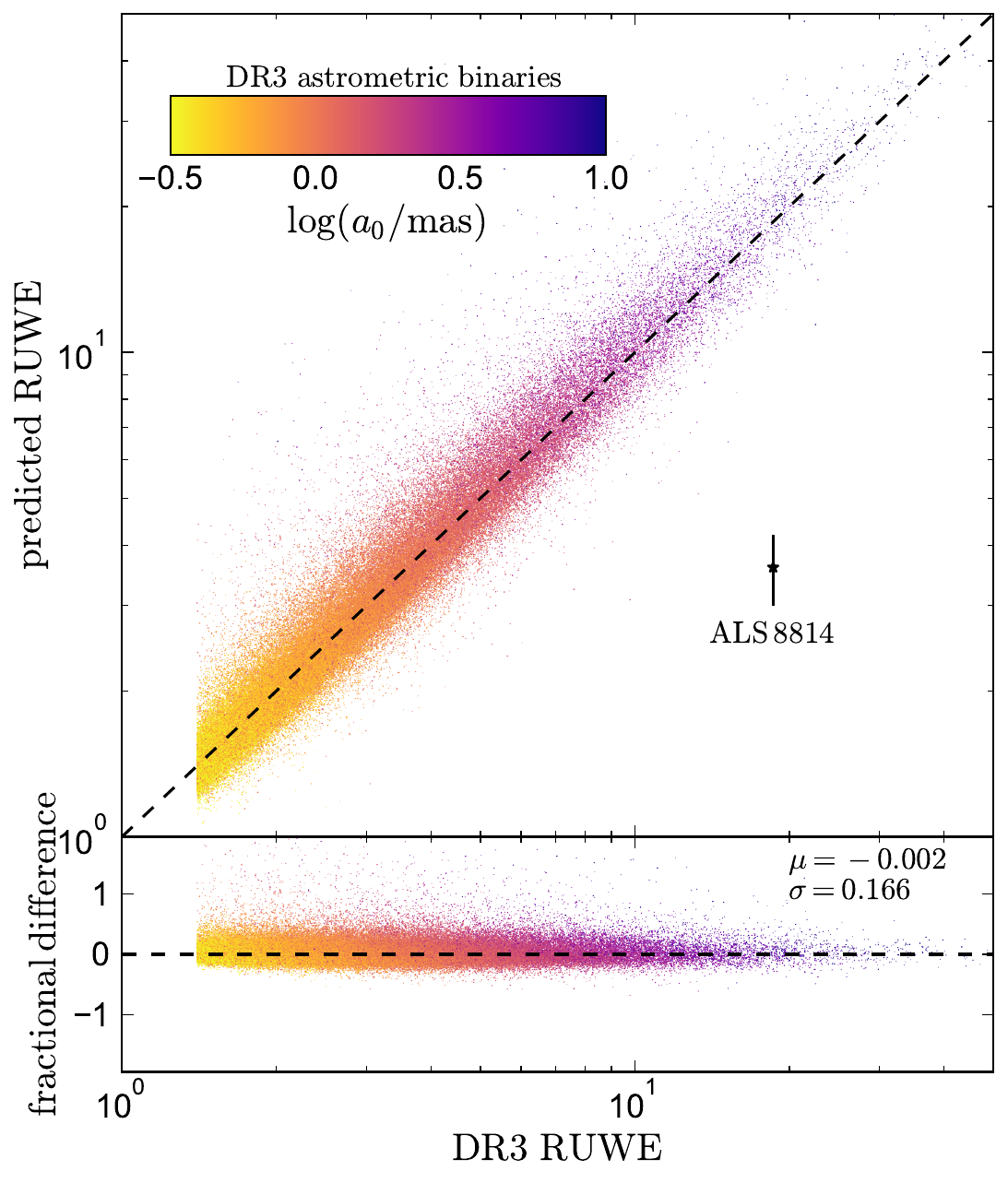}
    \caption{Observed and predicted \texttt{RUWE} of {\it Gaia} DR3 astrometric binaries (colored points) and ALS 8814 (black star). ALS 8814 has a much larger \texttt{RUWE} than predicted from its orbit for any plausible inclination or orientation. In contrast, the predicted \texttt{RUWE} of the DR3 astrometric binaries is fairly accurate. The simplest explanation is that a marginally-resolved tertiary or background star in ALS 8814 perturbs the astrometry of the binary. }
    \label{fig:ruwe}
\end{figure}

Figure~\ref{fig:ruwe} compares the observed and predicted \texttt{RUWE} of ALS 8814 to the astrometric binaries with published orbital solutions from {\it Gaia} DR3. \texttt{gaiamock} reliably predicts the \texttt{RUWE} values of these sources within about 20\%. ALS 8814 has a \texttt{RUWE} value more discrepant from the predicted value than any of these binaries, meaning that its \texttt{RUWE} is too large to be a consequence of binary motion alone. The fact that the binary likely contains a luminous secondary only exacerbates this discrepancy, since a luminous secondary will produce a smaller photocenter orbit. The observed \texttt{RUWE} is also too large to be a result of a marginally resolved Be star disk \citep[e.g.][]{Dodd2024} or photometric variability. 

The large \texttt{RUWE} of ALS 8814 is most likely a consequence of a third luminous source, either physically associated to ALS 8814 or in the background. As described by \citet{Lindegren2022}, marginally resolved companions with separations of order 0.1 arcsec -- which in most cases do not cause  measurable acceleration on {\it Gaia}-observable timescales, but still affect astrometry via blending -- lead to biased astrometry and large \texttt{RUWE}. \citet{Tokovinin2023} validate this prediction empirically, demonstrating that many sources with large \texttt{RUWE} have unresolved companions at separations of order 0.1 arcsec. Using \texttt{gaiamock}, we find that at an angular separation of 0.1-0.3 arcsec, a nearby source with flux ratio $\gtrsim 3\%$ could explain the observed \texttt{RUWE}. Such a source could likely be detected with high-resolution imaging or interferometry. 

\section{Nature of the system}
\label{sec:discussion}

\subsection{Could the companion be an artifact of V/R variations?}
\label{sec:VR_variations}

\label{sec:params}
\begin{figure*}
    \centering
    \includegraphics[width=\textwidth]{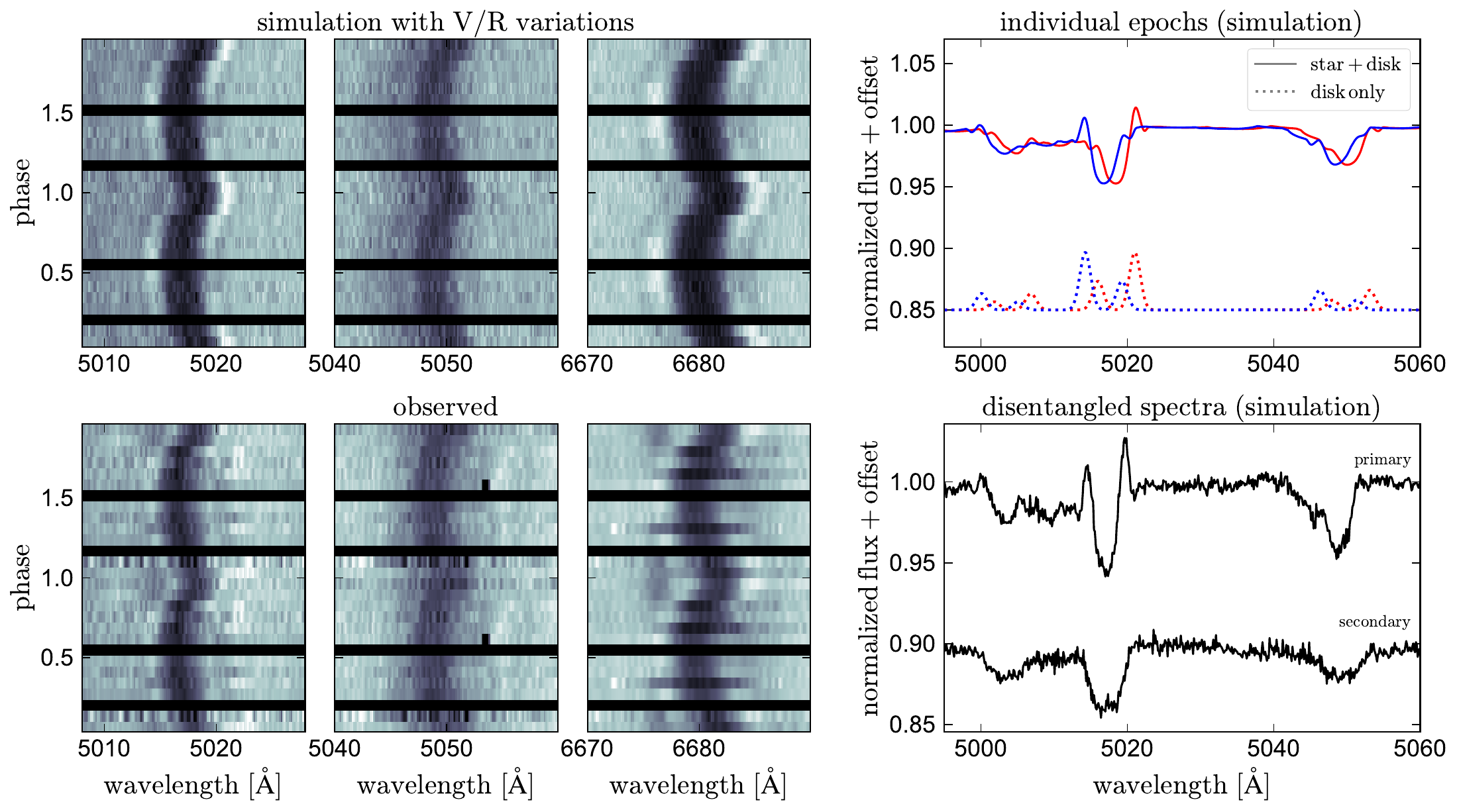}
    \caption{Effects of V/R variations on spectral disentangling. We simulate spectra of a binary with a single luminous star surrounded by a disk that undergoes phase-locked V/R variations (Equation~\ref{eq:f}). Spectra at opposite RV extrema are shown in the upper right panel; upper left panels show all the trailed simulated spectra. Lower right panel shows the results of applying spectral disentangling to the simulated spectra. Although no luminous companion was assumed in generating the data, disentangling produces a secondary spectrum that resembles the secondary extracted from the observed data (Figure~\ref{fig:vsini}). The bottom panel shows the observed spectra of ALS 8814 for comparison. There are two key differences between these data and the simulation: (a) the data do not show the coherent V/R variations that cause the spurious secondary detection in the simulations, and (b) the anti-phase motion of a second set of absorption lines seen in the data is not present in the simulations. We conclude that V/R variations are unlikely to explain the evidence for a secondary in the data. }
    \label{fig:VR_variations}
\end{figure*}

Many Be star binaries exhibit phase-locked V/R variations, meaning that the shape of their double-peaked emission lines varies coherently on the orbital period \citep[e.g.][]{Miroshnichenko2023}. Such variations violate the basic assumption of spectral disentangling -- that the component spectra are time-invariant -- and thus can be expected to introduce spurious features in the recovered spectrum of the secondary. Here we investigate whether V/R variations could explain the various lines of evidence for a luminous secondary in ALS 8814. 

We simulated phase-locked V/R variations in a binary with a single luminous component and apply disentangling to the resulting spectra. We model each double-peaked emission line with a sum of two Gaussians separated in velocity space by $\delta v = 300\,{\rm km\,s^{-1}}$. We set the width of each peak to a FWHM of $100\,{\rm km\,s^{-1}}$, and we adjust the mean amplitude of each line to match the observed spectra. We simulate such double-peaked lines at wavelengths of 5001.5, 5015.68, 5047.74, and 6678.15\,\AA, with mean amplitudes of 0.01, 0.035, 0.01, and 0.045 relative to the normalized spectra. We assume that the emission lines move coherently with a single luminous primary, which we model with a TLUSTY model with $T_{\rm eff} = 26\,{\rm kK}$, $\log g=4.0$, and $v \sin i = 250\,{\rm km\,s^{-1}}$ following \citet{An2025}. We chose the separation and width of the lines to best match the observed spectra.

In all known Be stars with well-characterized phase-locked V/R variations, the sign of the variation is such that the red peak of the emission lines is strongest when the Be star is at maximum blueshift, while the violet peak is strongest when it is at maximum redshift \citep{Miroshnichenko2023}. We find that simulated V/R variations with this phasing produce a disentangled secondary spectrum with emission lines that does not resemble the results of disentangling the observed spectra. 

Therefore, we explored a scenario where V/R variations occur with sign opposite to that usually observed. We introduced modulation factors for the violet and red peaks of the emission lines, $f_{\rm V}$ and $f_{\rm R}$: 

\begin{align}
    \label{eq:f}
    f_{{\rm V}}&=1-\frac{1}{3}\sin\left[2\pi\left(\phi-\frac{1}{4}\right)\right] \\ f_{{\rm R}}&=1+\frac{1}{3}\sin\left[2\pi\left(\phi-\frac{1}{4}\right)\right],
\end{align}
which vary between $2/3$ and $4/3$. At each epoch, we multiply the amplitude of the violet and red emission peaks by their respective modulation factors. The resulting spectra have stronger red peaks at maximum redshift, and stronger violet peaks at maximum blueshift. 

The results of this experiment are shown in Figure~\ref{fig:VR_variations}. In the presence of V/R variations with the amplitude and direction described by Equation~\ref{eq:f}, disentangling does produce a secondary spectrum with broad absorption lines, resembling the secondary we recover from the real data. This occurs because the simulated V/R variations are nearly equivalent to a fixed absorption component of amplitude 2/3 the mean amplitude of each emission line, which moves on top of the emission line and is close to stationary in binary's center of mass frame.

However, the simulated epoch spectra from this experiment do not resemble the actual spectra (bottom left panel of Figure~\ref{fig:VR_variations}) in detail. Most importantly, the simulated spectra with V/R variations show no signs of a second set of absorption lines moving in anti-phase with the primary, as is seen in the observed data. Moreover, the strength of any coherent V/R variations in the data is significantly weaker and less coherent than what is required in the simulations to produce spurious detection of a secondary. We also find that the simulated data with V/R variations are best-fit by a stationary secondary ($K_2=0\,{\rm km\,s^{-1}}$), while the observed data are better fit by $K_2 > 0\,{\rm km\,s^{-1}}$. 

Given the lack of strong phase-locked V/R variations apparent in the data, we also tested the effects of V/R variations that are uncorrelated with orbital phase. To this end, we draw $f_{\rm V}$ from a uniform distribution, $\mathcal{U}(2/3, 4/3)$ and set $f_{\rm R}=2-f_{\rm V}$. We find that such random V/R variations do not lead to the detection of a spurious secondary in disentangling. 

V/R variations are also unlikely to explain the apparent stationarity of the strong absorption lines (Figure~\ref{fig:lowres}), since the emission lines in the higher-order Balmer lines are weaker and narrower than the absorption lines. In the simulations described above, V/R variations lead to obvious asymmetries in the line cores, with Doppler shifts still visible in the wings, contrary to what is observed. Emission lines with fine-tuned shapes and V/R variations could in principle exactly cancel the motion of absorption lines. However, emission features will become progressively weaker in higher-order Balmer lines, and none of the higher-order lines in the LRS spectra display the predicted asymmetries or Doppler shifts. We conclude that while V/R variations may lead to biases in the secondary spectrum inferred from disentangling -- and thus, further data are required to robustly determine the binary parameters -- they are unlikely to explain the several independent lines of evidence for a luminous secondary.

\subsection{SB2 or static tertiary?}
\label{sec:lum_tertiary}

\begin{figure*}
    \centering
    \includegraphics[width=\textwidth]{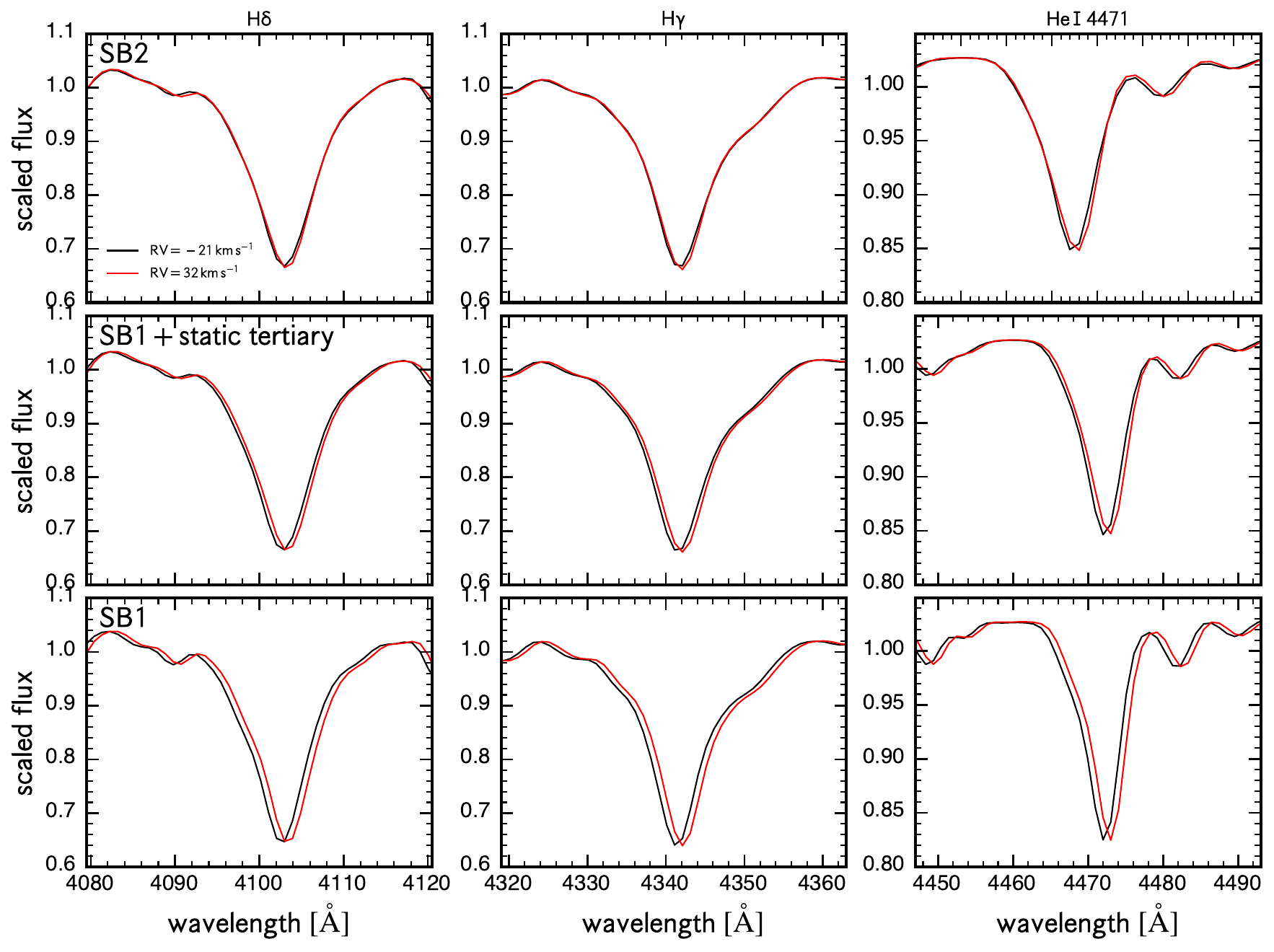}
    \caption{Simulation of the LRS spectra shown in Figure~\ref{fig:lowres} for three different physical scenarios. Top row shows two luminous and equally-bright stars moving in anti-phase. Middle row shows one luminous star moving, while another luminous component -- presumed to be a distant tertiary -- is stationary. Bottom row shows a single luminous star orbiting a dark companion. Only the case of a luminous binary (top row) can reproduce the lack of RV shifts between epochs that is observed in the data. }
    \label{fig:lrs_sim}
\end{figure*}

The astrometric evidence for a marginally-resolved tertiary, together with the rather weak constraints on $K_2$ from spectral disentangling, raise the question whether the luminous companion visible in the spectra might actually be a distant tertiary. Such a tertiary would have a sufficiently long period that its motion would not be detectable during the few-year baseline of the LAMOST observations, and its RV would be within a few $\rm km\,s^{-1}$ of the binary's center-of-mass RV.

To explore this possibility, we simulated the LRS spectra near the extrema of the observed spectra shown in Figure~\ref{fig:lowres}. We consider three possible physical scenarios: (1) the spectra are dominated by two luminous stars moving in anti-phase with equal RV amplitudes, (2) only the primary is moving, and the second luminous source is a static tertiary, and (3) the primary is the only luminous source contributing to the spectra. In all cases, we assume $T_{\rm eff} = 26\,{\rm kK}$ and $\log g = 4$ for both components, $v\sin i =130\,{\rm km\,s^{-1}}$ for the primary, and $v\sin i=250\,{\rm km\,s^{-1}}$ for the secondary, and a flux ratio of 0.5. We note that the primary's true $v\sin i$ is likely larger than $130\,{\rm km\,s^{-1}}$, but this value produces absorption line cores with shape similar to the combined primary + disk spectrum (Figure~\ref{fig:vsini}).

The results of this experiment are shown Figure~\ref{fig:lrs_sim}. The SB1 case (bottom panel; i.e., only one luminous source) can be immediately dismissed, because it predicts quite obvious and coherent shifts in all the strong lines, inconsistent with observations. The SB1 + static tertiary scenario (middle panel) predicts shifts that are weaker by about a factor of two, since the primary's motion is diluted by an unmoving tertiary contributing half the light. However, the predicted shifts between the two epochs are still manifestly stronger than those in the observed data, disfavoring this scenario. Finally, the ``SB2'' case with two equally-bright stars moving in anti-phase predicts behavior very similar to what is observed: the Balmer lines appear basically static between epochs, while the \spectral{He}{I}{4471} line shifts slightly redward in its core but not in its wings.

We conclude that a static tertiary cannot explain the lack of RV variability seen in the strong absorption lines. This does not mean that no tertiary exists, but rather that if it does, it does not contribute much to the total flux, and a luminous close companion must also be present. 

\subsection{Evolutionary state}
\label{sec:classical}

ALS 8814 does not appear to contain a BH. However, the system remains unusual in several ways. First, the companion appears to be a normal main-sequence star, yet main-sequence companions to Be stars are rare \citep{Bodensteiner2020}, presumably because many Be stars are mass transfer products. Second, the Be star appears to have the largest RV semi-amplitude and one of the highest mass functions of known classical Be stars, likely again reflecting the fact that most Be star companions are stripped products of mass transfer. Third, the orbit is significantly eccentric, while most (but not all) Be stars with stripped companions have nearly circular orbits \citep{Muller-Horn2025}.

These considerations and the likely presence of a distant tertiary suggest the possibility of an unusual formation history for ALS 8814. On the other hand, a few other Be stars are known with orbital configurations that do not suggest a history of accretion \citep[e.g.][]{Klement2021, El-Badry2022_ngc2004, Kervella2022, Rast2024}, so it is also possible that the Be star was simply born spinning rapidly and reached near-critical velocity through single-star evolutionary processes.  A more precise measurement of the component parameters and the dynamical mass ratio will be required to better constrain the system's evolutionary state and formation history and determine whether it is a in a pre- or post-interaction stage.

\section{Conclusions}
\label{sec:conclusions}
We have reanalyzed the LAMOST spectra of ALS 8814, an emission-line binary recently proposed by \citet{An2025} to contain a quiescent BH orbited by a Be star. Our most important finding is that there is evidence for at least two luminous components contributing roughly equally to the optical spectra and moving in anti-phase, removing the need for a BH. The system nevertheless remains unusual among Be stars and may represent an unusual formation history or short-lived evolutionary phase. Our main results are as follows. 

\begin{enumerate}
    \item {\it Spectral variability}: the line profiles of most absorption lines in ALS 8814 are not static, but vary with orbital phase. Inspection of the system's trailed spectra reveals two sets of absorption lines moving in anti-phase (Figure~\ref{fig:trailed}), implying the presence of two luminous objects orbiting one another. 
    \item {\it Inconsistent Doppler shifts of emission and absorption lines}: While the emission lines in ALS 8814 are strongly RV variable, the strong absorption lines vary significantly less (Figure~\ref{fig:lowres}), implying that emission and absorption lines trace different objects. We interpret this as evidence that the emission lines entirely trace the primary, but about half the light in the absorption lines traces the secondary, which moves in the opposite direction of the primary and cancels out the lines' average RV shifts. 
    \item {\it Spectral disentangling reveals a luminous companion}: We used three different spectral disentangling methods to disentangle the lines of the two components. This yields a primary with emission lines and comparatively narrow absorption lines, and a secondary with broad absorption lines (Figures~\ref{fig:disentangled} and~\ref{fig:comparison}). The disentangled spectra both resemble spectral models for stellar photospheres (Figure~\ref{fig:vsini}), with a best-fit flux ratio of about 0.5. Both components likely have high projected rotation velocities of order $300\,{\rm km\,s^{-1}}$, but the presence of emission lines prevents a precise measurement of $v\sin i$.

    We also performed disentangling on simulated spectra to assess whether the evidence for a luminous secondary could be an artifact of emission line variability (Figure~\ref{fig:VR_variations}). We find that while phase-locked variability can indeed lead to detection of a spurious secondary, the observed spectra do not display the type of variability required to explain the disentangling results with a single luminous component. Moreover, emission line variability alone is unlikely to explain the stationarity of the strong absorption lines.

    \item {\it Large astrometric excess noise}: ALS 8814 has very significant astrometric excess noise, as quantified by the \texttt{RUWE} parameter in {\it Gaia} DR3. The source's \texttt{RUWE} is too large to be explained by orbital motion for any plausible orbit and distance (Figure~\ref{fig:ruwe}). The simplest explanation is that ALS 8814 is a hierarchical triple, with a marginally-resolved tertiary at a separation of order 100\,au that disturbs the astrometry of the inner binary.
    \item {\it The luminous companion must be RV-variable}: Although a luminous and static tertiary could contribute some of the light in ALS 8814, it cannot be the only companion contributing to the spectra. In order to explain the lack of apparent RV variability in the system's strong absorption lines (Figure~\ref{fig:lowres}), a source that is comparably bright to the primary must move in the opposite direction to it (Figure~\ref{fig:lrs_sim}). 
\end{enumerate}


\section*{acknowledgments}
We thank the referee for a constructive report and Yang Huang and Meng Sun for useful discussions. This research was supported in part by the CCA at the Flatiron Institute who provided hospitality during the Stable Mass Transfer 2.0 Workshop.

This research was supported by NSF grant AST-2307232 and by a Sloan research fellowship. TS acknowledges funding from the European Research Council (ERC) under the European Union’s Horizon 2020 and Horizon
Europe research and innovation programme (grant agreement number 101164755: METAL) and the Israel Science Foundation
(ISF) under grant number 0603225041.
This research is supported by the the Flemish Government under the long-term structural Methusalem funding program, project SOUL: Stellar evolution in full glory, grant METH/24/012 at KU Leuven. It is also supported by from the Research Foundation – Flanders (FWO) (grant agreement G0ABL24N) and from the KU Leuven Research Council through grant iBOF/21/084.
\newpage

\newpage
\bibliographystyle{mnras}

\end{document}